\documentclass[systeme]{compas2020}
\usepackage{url}
\toappear{1} 

\begin{document}

\title{Infrastructure de Services Cloud FaaS sur n\oe uds IoT}
\shorttitle{PyCloudIoT: FaaS over IoT nodes}

\author{David Fernández Blanco, Frédéric Le Mouël}%
\address{Univ Lyon, INSA LYON, CITI, F-69621 Villeurbanne, France\\
david.fernandez-blanco@insa-lyon.fr et  frederic.le-mouel@insa-lyon.fr}

\date{\today}

\maketitle
\begin{abstract}
  Dans cet article, nous décrivons l'infrastructure cloud PyCloudIoT. PyCloudIoT s'appuie sur un modèle de cloud computing FaaS pour du calcul numérique déporté vers une ferme de calcul composée de noeuds à capacités restreintes au lieu d'un datacentre puissant. Cette infrastructure vise à tirer profit des ressources inutilisées déployées sur n\oe uds IoT sans augmenter significativement leur consommation d'énergie et, en même temps, à rapprocher ces ressources des utilisateurs pour réduire la latence, en développant un modèle cloud en bord de réseau.
 
  \MotsCles{Distributed Systems, Cloud Computing, FaaS, Edge Computing, IoT, Python}
\end{abstract}
\vspace{7px}
Selon \textit{statista} \cite{statista}, depuis la sortie des objets connectés en 2009, le nombre d'appareils et le flux de données ont augmenté de façon exponentielle, allant de 1 milliard en 2009 jusqu'à plus de 20 milliards en 2020 et prévoyant de doubler au cours des deux prochaines années. En outre, les données sont créées plus près des utilisateurs, de manière plus rapide, plus hétérogène et en quantité plus élevée qu'auparavant. De plus, les données gagnent de la valeur quand elles sont analysés au fur et à mesure que produites, ce qui entraîne une explosion de la charge réseau. Ce besoin de délai le meilleur possible et de traitement de grands volumes de données posent des problèmes pour les architectures basées sur des serveurs centralisés - comme les architectures en nuages distantes.

\noindent Dans le même temps, les ressources des périphériques IoT sont très souvent sous-exploitées. Cependant, ces périphériques sont plus susceptibles de rencontrer des pannes - liées au réseau ou au processeur ou à la limite de la batterie, ce qui limite leur fiabilité et la charge de calcul qu'ils peuvent assumer. Nous proposons PyCloudIoT - infrastructure cloud FaaS de calcul numérique en Python, permettant l'utilisation de ces ressources inexploitées. L'objectif est de raccourcir la distance utilisateur-serveur afin de réduire le stress du réseau et la latence en développant un modèle de cloud computing FaaS (\textit{Fonction as a Service}) dans lequel la ferme de calcul est composée de n\oe uds IoT placés à la périphérie du réseau. Nous cherchons à maximiser le compromis entre performance, énergie consommée et tolérance aux pannes pour s'adapter à cet environnement IoT contraint et sujet aux pannes.

\section{État de l'art}

Dans cette section, nous présentons les approches ayant des objectifs similaires: faire du calcul distribué sur des environnements très limités, même si les capacités ne sont pas toujours aussi restreintes dans toutes les approches. 

Si nous regardons les approches classiques de cloud computing, basés sur ferme de serveurs et non sur IoT, nous pouvons trouver des systèmes tels que \emph{PyWren} \cite{pywren} proposant une architecture de stateless-cloud cloud élastique et passant à l'échelle. Leur plate-forme a pour objectif premier une prise en main facile, d'avoir un temps d'exécution réduit, un planificateur de tâches rapide, l'équilibrage de charge et la possibilité d'ajouter des dépendances au run-time. Ils ont également implémenté la ré-exécution symétrique des tâches pour renforcer la tolérance aux pannes du système et plusieurs optimisations sur le parallélisme du code. Aucun aspect de consommation énergétique n'y est toutefois abordé.

Dans l'article \emph{Middleware Implementation in Cloud-MANET Mobility Model for Internet of Smart Devices } \cite{MANET}, nous pouvons voir une intégration du Cloud et du réseau MANET. Cette approche utilise le réseau ad-hoc MANET pour connecter des appareils intelligents entre eux puis implémente une couche android middleware afin d'offrir un service de cloud computing aux utilisateurs. Cependant, les capacités des appareils utilisés, même étant un environnement constraint, sont importantes (CPU et GPU complexes), et l'optimisation finale porte uniquement sur l'amélioration de distance de couverture de bord de réseau et le débit de celui-ci.

Sur un autre périmètre, celui des réseaux véhiculaires, nous pouvons trouver d'autres approches, comme \emph{VANET-cloud: a generic cloud computing model for vehicular Ad Hoc networks} \cite{VANET}, qui implémentent du cloud computing sur réseaux véhiculaires afin de réduire le délai, et l'énergie consommée, augmenter l'efficacité, l'évolutivité, la fiabilité et la sécurité du cloud computing pour les systèmes de transport intelligents (STI). Ce système intègre de nouvelles ressources embarquées déjà présentes sur les véhicules pour étendre l'environnement cloud traditionnel. La fiabilité envisagée reste toutefois sur la communication inter-véhiculaire et non sur d'autres types de pannes.

 \emph{cuCloud} \cite{cuCloud} propose un système partageant notre logique: créer un réseau ad-hoc opportuniste avec nos propres appareils, PC dans leur cas, pour améliorer les capacités de calcul. Original, basé sur une participation volontaire de machines aux ressources sous-utilisées, \emph{cuCloud} se concentre uniquement sur le passage à l'échelle et les performances, sans prise en compte de l'énergie. Ce système est de plus basé sur la réplication des machines virtuelles et non sur l'exécution de fonctions légères comme nous le proposons.

\section{PyCloudIoT : un infrastructure Cloud FaaS pour calcul numérique sur n\oe uds IoT}
Dans cette section, nous commencerons par décrire les messages et canaux ainsi que les grands traits de l'architecture de notre modèle. Dans la deuxième partie, nous présenterons la stratégie de consensus mise en \oe uvre et finalement, nous présenterons les techniques de partitionnement des graphes implantées dans l'architecture.

\subsection{Architecture et messages}
\begin{figure}[h!]
    \centering
    \includegraphics[width =0.60\textwidth]{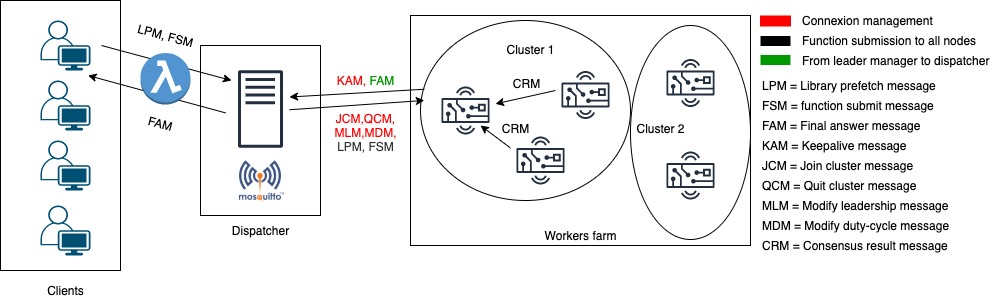}
    \caption{Acteurs de PyCloudIoT}
    \label{fig:image_mess}
\end{figure}

\begin{figure}[h!]
    \centering
    \includegraphics[width = 0.6\textwidth]{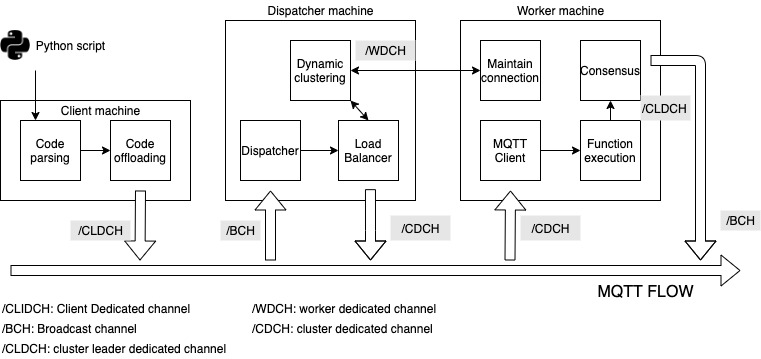}
    \caption{Fonctionnalités des acteurs de PyCloudIoT}
    \label{fig:my_label2}
\end{figure}

Comme décrit dans les figures \ref{fig:image_mess} et \ref{fig:my_label2}, PyCloudIoT est composé de trois types de n\oe uds: les workers, les dispatchers et les clients.

\textit{\textbf{Les clients:}} un client soumet des fonction Python à exécuter - fonctions annotées suite à l'analyse syntaxique du langage - via l'API PyCloudIoT installée localement.
    
\textit{\textbf{Le dispatcher:}} le dispatcher organise l'exécution des fonctions du client en asynchrone dans l'ensemble des groupes de Workers. Ce n\oe ud est en charge du regroupement dynamique (\emph{clustering} - expliqué dans la section \ref{Graph partitioning}), de l'exécution des tâches (\emph{offloading}) entre les différents groupes de Workers et de l'équilibrage des charges.
    
\textit{\textbf{Les workers:}} ces n\oe uds contiennent un processus qui se connecte de manière asynchrone au Dispatcher. Ils sont attribués à un groupe où un consensus à leader unique se chargera d'exécuter en parallèle les tâches .

\subsection{Description de l'algorithme de consensus} \label{consensus}
L'algorithme de consensus est basé sur Raft, avec des modifications pour l'adapter aux rôles et à la logique de notre infrastructure. Dans un premier temps, nous expliquons comment notre algorithme de consensus gère les interactions entre les Workers et Leaders. Pour commencer, un leader est considéré comme indisponible lorsque le Broker ne reçoit pas de message Keepalive pendant plus de 30 secondes. En deuxième lieu, l'élection des leaders - un pour chaque cluster de Workers - a lieu. Lorsqu'un leader est perdu ou lorsqu'un nouveau n\oe ud avec un duty-cycle plus performant rejoint le cluster, l'élection du leader a lieu. Alors qu'en Raft ce processus se déroule entre les n\oe uds du cluster avec un mechanisme de vote, dans notre algorithme, ceci a lieu sur le Dispatcher. Pour cela, le cluster prendra en compte 2 arguments pour déterminer le prochain leader : la périodicité du duty-cycle et le moment où la prochaine connexion du n\oe ud aura lieu. La première étape de l'élection du leader consiste à examiner les n\oe uds actifs du cluster et à les ordonner par périodicité de duty-cycle, nous privilégions les n\oe uds avec la périodicité la plus rapide. Ensuite, nous regarderons le prochain timestamp de connexion (calculé en ajoutant le dernier timestamp de connexion à la périodicité). Parmi les n\oe uds de même périodicité, celui qui se reconnecte le plus tôt sera choisi comme leader.  
Enfin, le n\oe ud que le Dispatcher notifie être le leader souscrit au canal dédié au leader du cluster et commence à traiter les messages dirigés vers le leader. La plus grande différence entre notre algorithme de consensus et Raft est que nous n'avons plus besoin que les n\oe uds du cluster connaissent qui est le leader puisque le canal du leader sera toujours le même.
Enfin, l'algorithme de consensus a une dernière étape lors de l'envoi d'une réponse à la demande d'un client. Ce mécanisme fait que le leader n'ait pas à attendre jusqu'à que toutes les réponses des n\oe uds du cluster lui soient arrivées, seulement jusqu'à ce qu'il reçoive une majorité de réponses avec le même résultat. 

\subsection{Techniques de tolérance aux pannes}
Grâce au mécanisme décrit dans la section précédente, PyCloudIoT est résistant aux défauts byzantins pour l'ensemble des n\oe uds IoT (si une tolérance aux pannes devait être mise en place pour les noeuds clients et Dispatcher, un algorithme classique de consensus peut leur convenir). Le fait d'avoir un processus de vote avant de donner une réponse finale le rend $ \frac{N-1}{2} $ tolérant. En plus des techniques de tolérance aux pannes liées au consensus, nous avons mis en \oe uvre d'autres mécanismes pour défendre le système contre d'autres défaillances actuelles sur les systèmes distribués \cite{failure-types}: défaillances réseau, avec l'ajout d'un protocole de message asynchrone MQTT. Le système est tout de même protégé contre les deux défaillances lors de l'exécution d'une tâche ou lorsqu'il est inactif grâce au concept de \textit{consume} présent dans MQTT. De plus, grâce au système Keep-alive mis en \oe uvre, lorsqu'un n\oe ud tombe en panne, il sera purgé de l'infrastructure après 30 secondes. Finalement, si le n\oe ud se bloque lors de l'exécution d'une tâche, nous aurons les autres n\oe uds du cluster pour couvrir l'échec (ayant besoin d'au moins un n\oe ud du cluster ne rencontrant pas de défaillance).

\subsection{Techniques de partitionnement de graphe} \label{Graph partitioning}
Dans une infrastructure IoT avec des périodes de veille (\textit{sleep-awake}) des n\oe u ds, le regroupement et partitionnement est indispensable pour maximiser les performances ainsi que la stabilité et la fiabilité de l'infrastructure. Nous avons abordé ce problème comme un problème classique de partitionnement de graphe (\textit{graph partitioning}), devant regrouper par 3 à 7 les n\oe uds disponibles, sur deux critères: la période de réveil (\textit{duty-cycle}) et l'heure de la dernière fois de communication avec le Dispatcher. L'algorithme de clustering suit cinq étapes simples. Pour commencer, le Dispatcher ordonne les appareils en fonction de leur duty-cycle, puis par le dernier timestamp de connexion reçu. Ensuite, il utilise le nombre de n\oe uds disponibles pour décider du nombre optimal de clusters entre 3 et 7 n\oe uds. Après cela, il prend le n\oe ud disponible le plus rapide et le place comme leader de chaque cluster requis. Enfin, il remplit les clusters en alternance avec le n\oe ud disponible le plus rapide et le plus lent.

\begin{figure}[h!]
    \centering
    \includegraphics[width = 0.6\linewidth ]{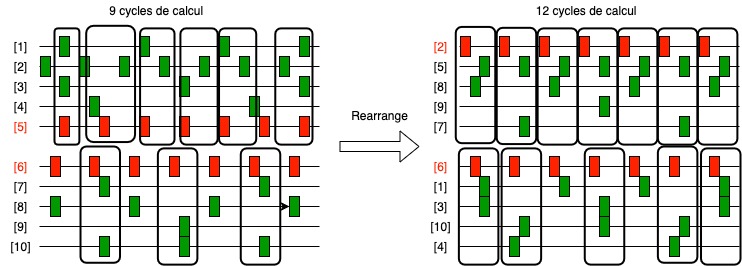}
    \caption{Exemple de repartitionnement du graphe de n\oe uds IoT en 2 clusters}
    \label{fig:my_label3}
\end{figure}

Comme détaillé sur la figure \ref{fig:my_label3}, après regroupement, le nombre de \textit{Keep-alive} pour chaque n\oe uds par duty-cycle sur chaque cluster se rapproche de celui du leader - n\oe uds 2et 6. De plus, si nous comptons les cycles de calcul, nous avons une augmentation globale de 3, ce qui représente une augmentation des performances de 33 \% sur la même durée sur cet exemple.

\section{Implémentation, résultats et discussion}
\textit{\textbf{Matériel et banc de test:}} Pour les tests, nous avons utilisé un Macbook Pro 15 comme client et Dispatcher, et 7 cartes ESP-32 comme Workers. Nous avons testé avec un banc de test public \textit{numpy} \footnote{\url{github.com/serge-sans-paille/numpy-benchmarks/tree/master/benchmarks}} de calcul numérique, adapté pour notre infrastructure. Ces fonctions ont été exécutées en ajoutant une série d'annotations pour les rendre compréhensibles pour le parser du client.

\textit{\textbf{Métriques:}}
Dans les tests, nous avons pris en considération nos trois principaux facteurs de notre modèle: la tolérance aux pannes, la performances et la consommation d'énergie. Pour la consommation d'énergie, nous avons utilisé un modèle décrit dans \cite{cons}, incluant le courant moyen en mode actif, le courant moyen en mode veille et le duty-cycle. Nous avons inferé la consommation des ESP-32 \cite{esp-approx} dans notre cas, avec seulement besoin du WIFI et des modules de carte de base, par $ I_ {active} = 0,260 A $ et $ I_ {sleep} = 2,5 * 10 ^ {- 6} A $ alors que l'ESP-32 est en mode hibernation.
\vspace{10px}
\begin{figure}[h!]
\centering
\begin{minipage}{.5\textwidth}
  \centering
  \includegraphics[width=0.6\linewidth]{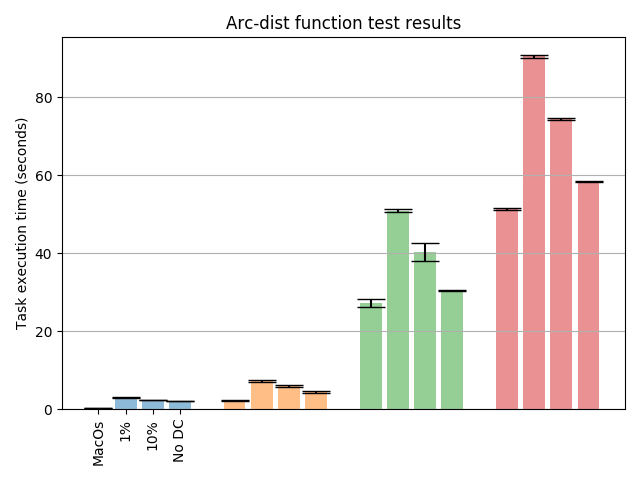}
\end{minipage}%
\begin{minipage}{.5\textwidth}
  \centering
  \includegraphics[width=0.6\linewidth]{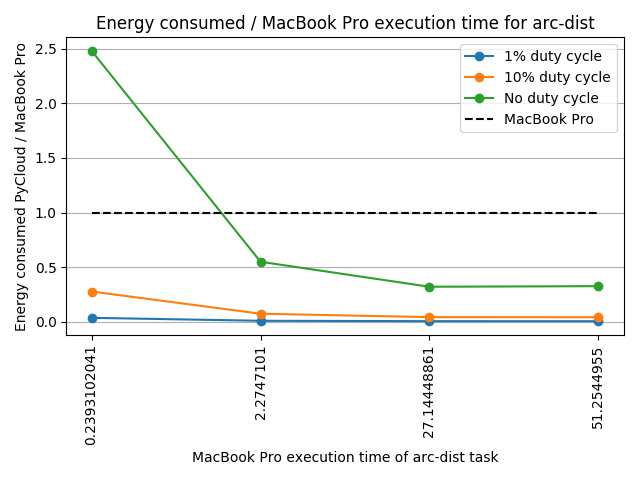}
\end{minipage}
  \caption{Temps d'exécution de la fonction arc-dist (pour 4 valeurs de plus en plus importantes de calcul) avec le comparatif entre PyCloudIoT (1\% duty-cycle, 10\% duty-cycle, pas de duty cycle) et MacBook Pro 15 (figure de gauche). Consommation énergétique de la fonction arc-dist (figure de droite).}
  \label{fig:perf-en}
\end{figure}

\begin{figure}[h!]
\centering
\begin{minipage}{.33\textwidth}
  \centering
  \includegraphics[width=\linewidth]{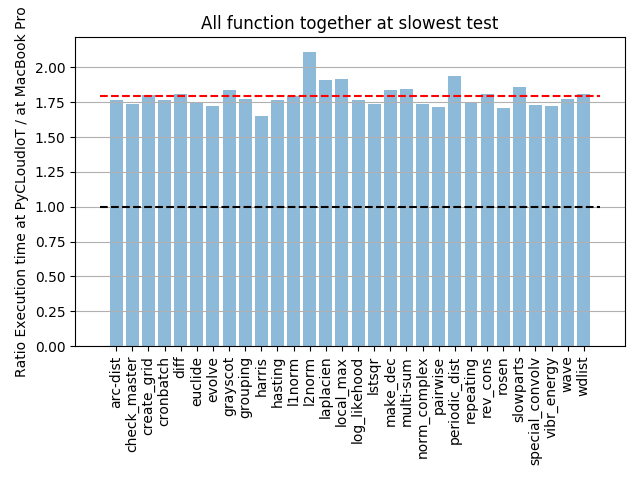}
\end{minipage}%
\begin{minipage}{.33\textwidth}
  \centering
  \includegraphics[width=\linewidth]{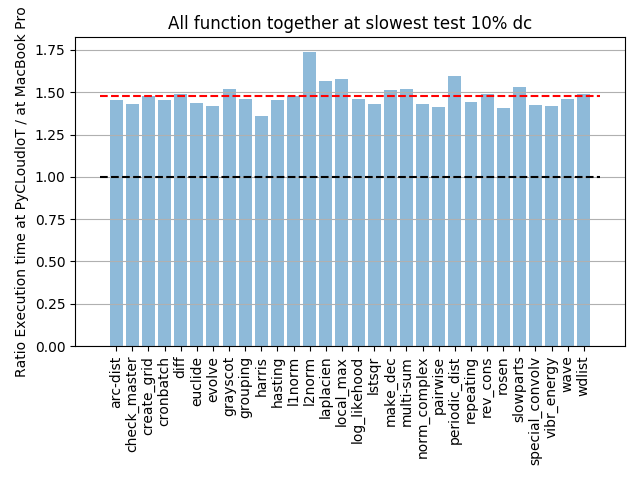}
\end{minipage}%
\begin{minipage}{.33\textwidth}
  \centering
  \includegraphics[width=\linewidth]{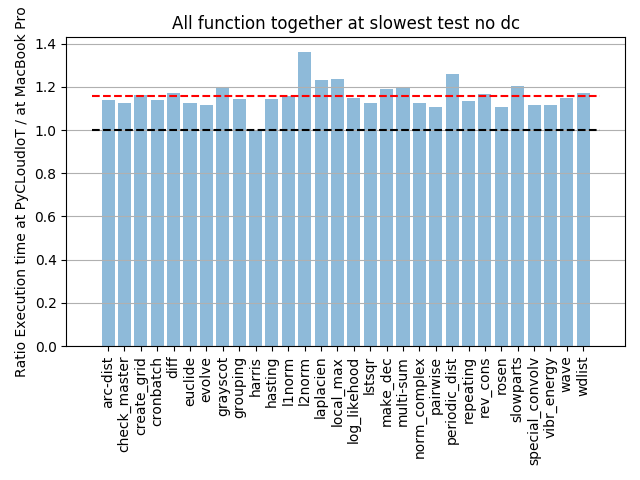}
\end{minipage}
  \caption{\small Rapport du temps d'exécution sur PyCloudIoT (1\% à gauche, 10\% au centre, pas de duty cycle à droite) / temps d'exécution sur MacBook Pro 15 pour toutes les fonctions du benchmark. (La ligne rouge représente la moyenne).}
  \label{fig:allto}
\end{figure}

Nous avons testé pour chaque fonction du benchmark numpy trois configurations (1\%, 10\% et sans duty-cyle) et en augmentant la complexité des calculs. Sur la figure \ref{fig:perf-en}, nous observons que le rapport entre le temps d'exécution du Macbook Pro et celui des différents configurations diminue avec l'exécution durée de la tâche convergeant vers environ 1,2 fois le temps d'exécution sans duty-cycle, 1,5 fois pour 10 \% duty-cycle et 1,8 fois pour 1 \% duty-cycle. Cependant, si nous observons la consommation énergétique, nous remarquons qu'il n'y a presque pas de différence entre le 10\% et le 1\% duty-cycle dans les tâches supérieures à 2 secondes. Le ratio performance/énergie présente donc le meilleur compromis pour 10\% duty-cycle.
Par ailleurs, sur la figure \ref{fig:allto}, nous pouvons remarquer que sur les tâches longues, nous retrouvons un comportement similaire pour toutes les fonctions testées, qui nous permet de trouver des seuils de convergence pour nos configurations. Cela montre que notre plate-forme se comporte essentiellement de la même manière pour toutes les fonctions du benchmark.

\begin{figure}[h!]
\centering
\begin{minipage}{.5\textwidth}
  \centering
  \includegraphics[width=0.7\linewidth]{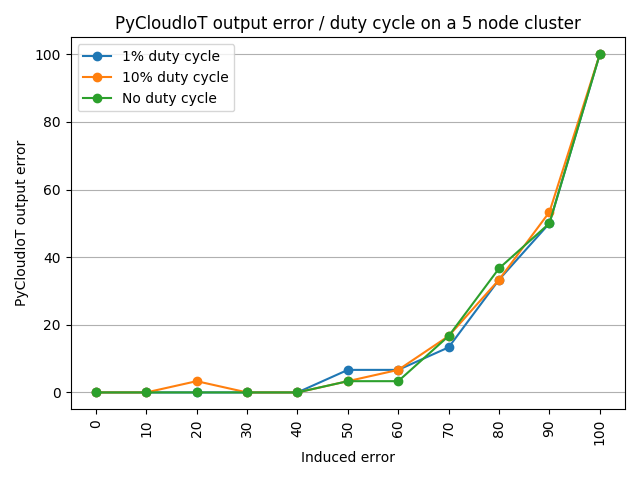}
\end{minipage}%
\begin{minipage}{.5\textwidth}
  \centering
  \includegraphics[width=0.7\linewidth]{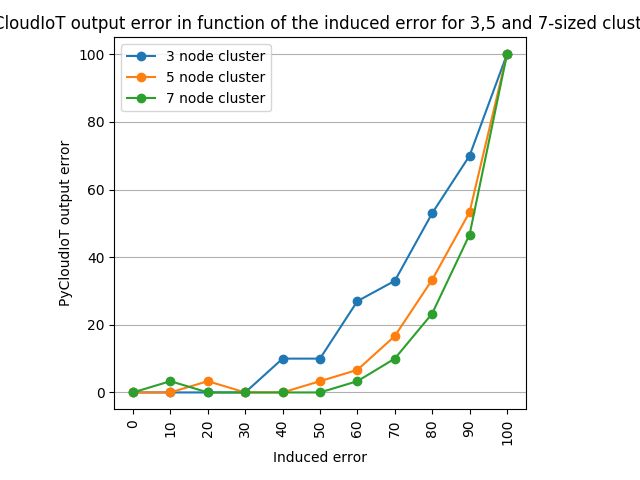}
\end{minipage}
  \caption{Taux d'erreur/erreurs induites en variant le duty-cycle (figure de gauche) puis la taille des clusters (figure de droite).}
  \label{fig:err}
\end{figure}

Enfin, si nous regardons l'influence du duty cycle sur la tolérance aux pannes, nous pouvons voir qu'il n'a aucune influence. Cependant, si nous nous intéressons à la taille des clusters, nous voyons que cela impacte directement la tolérance aux pannes du système, le rendant plus forte sur les clusters plus gros. En observant la figure de droite, le meilleur compromis panne/performance se retrouve avec 5 n\oe uds par cluster.
\vspace{0px}
\section{Conclusion et travaux futurs}

Après les tests, nous pouvons conclure que PyCloudIoT maximise le compromis cherché pour une configuration de cluster à 5 n\oe uds avec un duty-cycle de 10\%. Nous constatons une convergence stable pour les tâches moyennes / longues (de 5 à 60s) d'environ 1,4 fois plus longue que celle exécutée sur le Macbook Pro tout en diminuant presque totalement la consommation d'énergie (98\%). Comme travaux futurs, nous testerons et comparerons différents algorithmes de consensus, ainsi qu'optimisations de partitionnement de graphes. Ensuite, nous nous concentrerons sur la réalisation d'un analyseur syntaxique apprenant qui décidera des tâches à décharger en fonction des exécutions précédentes.

\vspace{-2pt}

\small \bibliography{main}

\def\No{\kern-.25em\lower.2ex\hbox{\char'27}}
\begin{thebibliography}{1}

\bibitem{MANET}
Alam (T.). --
\newblock {M}iddleware {I}mplementation in {C}loud-{MANET} {M}obility {M}odel
  for {I}nternet of {S}mart {D}evices. {\em IJCSNS}, vol.~17, n\No{} 5, mai
  2017, pp. 86--94.

\bibitem{VANET}
Bitam (S.), Mellouk (A.) et Zeadally (S.). --
\newblock {VANET}-{C}loud: a {G}eneric {C}loud {C}omputing {M}odel for
  {V}ehicular {A}d {H}oc {N}etworks. {\em IEEE Wireless Communications},
  vol.~22, n\No{} 2, f\'evrier 2015, pp. 96 -- 102.

\bibitem{pywren}
Jonas (E.), Pu (Q.), Venkataraman (S.), Stoica (I.) et Recht (B.~H.). --
\newblock {O}ccupy the {C}loud: {D}istributed {C}omputing for the 99\%. --
\newblock In {\em Proc. of SoCC'17}, pp. 445 -- 451, 2017.

\bibitem{esp-approx}
L.M.E. --
\newblock {I}nsight {I}nto {ESP32} {S}leep {M}odes \& {T}heir {P}ower
  {C}onsumption (online), 2020.

\bibitem{cuCloud}
Mengistu (T.~M.), Alahmadi (A.~M.), Alsenani (Y.), Albuali (A.) et Che (D.). --
\newblock {cuCloud}: {V}olunteer {C}omputing as a {S}ervice ({VCaaS}) {S}ystem.
  --
\newblock In {\em Proc. of CLOUD'18}, pp. 251 -- 264, 2018.

\bibitem{statista}
O'Dea (S.). --
\newblock Iot device installed base worldwide 2009-2020, 2020. Statista.

\bibitem{failure-types}
Sayson (M.). --
\newblock Types of failures in distributes systems, 2017. Notes from UBC CPSC
  416: Distributed Systems, University of British Columbia, Canada.

\bibitem{cons}
Zhu (T.), Zhong (Z.), Gu (Y.), He (T.) et Zhang (Z.-L.). --
\newblock Leakage-aware energy synchronization for wireless sensor networks. --
\newblock In {\em Proc. of MobiSys'09}, pp. 319 -- 332, juin 2009.

\end{thebibliography}

\end{document}